# Application of Ionizing Radiation for the Enhancement of Hydrogen Yield in Electrolytic Water Decomposition

**W. Ulmer[1]**
**Gesellschaft für Qualitätssicherung in der Medizin, Deizisau, Germany**
**E-mail: waldemar.ulmer@gmx.net**

**Abstract**
**During the past decay many experimental configurations to improve the yield of hydrogen by electrolysis. These attempts include tests of different materials for the electrodes. This study proposes the use of γ-radiation of waste sources of nuclear reactors. The related experimental configuration can be reduced to boxes filled with water exposed to ionizing radiation and capacitors as electrodes as well-known in other technical disciplies. Test measurements based on radiation of a linear accelerator (6 MV beam) provided significant differences of the yield of hydrogen between with and without radiation.**

## 1. Introduction

### 1.1. Description of the electrolytic process

The production of hydrogen ($H_2$) using electrolyzers is becoming of growing importance, since, on the one hand, the reserves of fossil fuel are decreasing, and, on the other hand, the combustion of these fossil fuels has evoked severe environmental problems through the release of carbon dioxide ($CO_2$) and other harmful gases. The principle of electrolytic water decomposition to gain $H_2$ and oxygen ($O_2$) using electric energy is an old method in electrochemistry. However, this field has attracted significant attention in recent decades, which has led to the commercialization of this technology. The principle of electrolytic water splitting is shown schematically in Figure 1.

Water decomposes at the anode into $O_2$ and protons ($H^+$) at the cathode. The electrons and protons migrate to the cathode through a membrane or external circuit, respectively, where the $H_2$ production reaction occurs. The entire reaction takes place according to the following reaction, eq. (1):

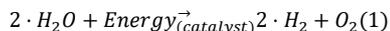

$$2 \cdot H_2O + Energy \xrightarrow{(catalyst)} 2 \cdot H_2 + O_2 \quad (1)$$

An external source of energy is required to drive the water splitting reaction. Theoretically, the minimum voltage for water splitting amounts to 1.23 V. In practice, however, higher voltages are needed to overcome kinetic limitations. The electric energy required to drive the water splitting reaction represents the major limitation of the large-scale production of $H_2$ via electrolysis. This is due to the significant cost for electricity that needs to be supplied to large electrolyzers, thus impeding their economic operation. In this invention, we present an apparatus that is capable to utilize ionizing radiation to reduce the amount of electric energy necessary to initiate the water splitting reaction. Hereby, ionizing radiation is introduced into a modified electrolyzer apparatus to decompose water. The radiation can originate, e. g. from the decay reactions of radioactive substances (e. g. nuclear waste), or naturally occurring radiation, can be used. The incorporation of the ionizing radiation into the electrolyzer apparatus effectively excites water molecules into excited states, which facilitates their decomposition into $H_2$ and $O_2$ gases at electrode surfaces. A ten-fold increase of the $H_2$ production has been measured in the presence of ionizing radiation relative to its absence.

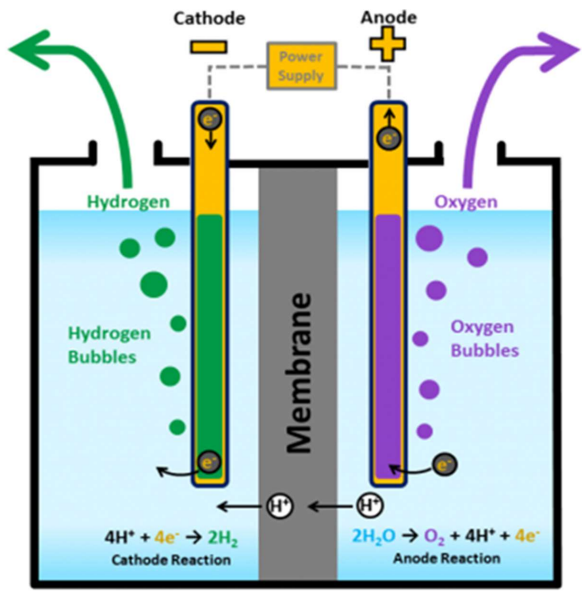

**Figure 1**. Schematic representation of the water electrolysis – the complete apparatus is referred to as electrolyzer (see e.g., Review in DOE Hydrogen Program, 2020, Kato, 2003, Walter, 2010).

## 1.2. Goal of the present study: Decay of radioactive waste as an alternative way as provider the energy for the electrolysis

The most important isotope in radioactive waste is $Cs^{137}$ with a half-time of 30 years, but due to the extreme activity various half-times are required until the intensity of the radiation emitted by this isotope is low enough so that it can be handled by usual technical methods. In order to avoid this long-time and unpredictable storage, there have been elaborated some concepts, which all can be referred to as *'transmutations'*. These concepts mainly deal with further irradiation of the decay products such as $Cs^{137}$ via fast neutrons in similar way as the nuclear reactors work with $U^{235}$ or similar appropriate nuclei. However, it is not clear how to apply the released energy and what is the amount of the energy to make the transmutations realistic (Ulmer, 2016). Besides the neutron path we may also think of irradiation with protons, since such installations are becoming rather wide-spread due to improvements in radiotherapy with protons. Two typical reactions of protons with $Cs^{137}$ are stated by eq. (2) and eq. (3).

$$p + Cs_{55}^{137} \rightarrow p + n + Cs_{55}^{136} + \gamma \quad (2)$$

$$p + Cs_{55}^{137} \rightarrow n + Ba_{56}^{137} + \quad \gamma \quad (3)$$

The two reactions with protons deserve particular interest, because according to eq. (2) another Cs-isotope is produced with a half-time of about 22 day in contrast to the 30 years of $Cs^{137}$ (Etspüler, 2012, Kulke, 2020). The keywords '*Radioactive waste – new exploitation technologies*' provide various references in the internet, a listing of them seems not to be necessary.

## 2. Methods

### 2.1.   Theoretical part – principles of interactions of radiation with water

Thermodynamically, the water splitting reaction is an uphill reaction. A minimum amount of 571.8 kJ mol[-1] must be supplied to drive this reaction according to eq. (4).

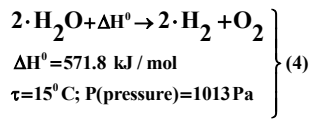

$$2 \cdot H_2O + \Delta H^\circ \rightarrow 2 \cdot H_2 + O_2$$
$$\Delta H^\circ = 571.8 \ kJ / mol$$
$$\tau = 15^\circ C; \ P(pressure) = 1013 \ Pa$$
$$(4)$$

This energy can be supplied in the form of electricity, light, or heat. When photons of sufficient energy level (h•ν) interact with water molecules, this can result in the formation of excited states, according to eq. (3):

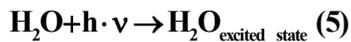

$$H_2O + h \cdot \nu \rightarrow H_2O_{excited \ state} \quad (5)$$

The nature of the excited state of the water molecule depends on the energy level of the incident photon ($E_{photon}$). The ionization energy of $H_2O$ amounts to 12 eV in the gaseous state, whereas in fluid form it mainly depends on the temperature and eventually on a further solvent, in general about 9 eV. If $E_{photon}$ exceeds the energy required to ionize the water molecule ($E_{ion}$), i. e. $E_{photon} > $ (or >>) $E_{ion}$, then an electron can be ejected from the water molecule. A water cation and a free electron are released according to eq. (4).

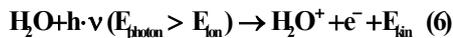

$$H_2O + h \cdot \nu \ (E_{photon} > E_{ion}) \rightarrow H_2O^+ + e^- + E_{kin} \quad (6)$$

This effect is referred to as the Einstein photo effect. If such a situation occurs, the collision of the released electron with $H_2O$-molecules in its environment can lead to the ionization of additional water molecules according to eq. (5):

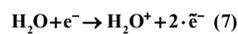

$$H_2O + e^- \rightarrow H_2O^+ + 2 \cdot \widetilde{e}^- \quad (7)$$

The notation $\widetilde{e}^-$ in eq. (5) indicates that the kinetic energy of the two released electrons is lower than that of the impinging electron $e^-$. If the photon energy $E_{photon} > 15$ keV or $>> 15$ keV a similar physical process occurs, known as Compton effect. Here, the energy level of the incident photon γ is so high that its collision with a water molecule causes the ejection of an electron $e^-$ and the additional ejection of a 'recoil' photon, γ', as shown in eq. (6).

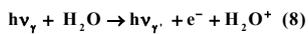

$$h\nu_\gamma + H_2O \rightarrow h\nu_{\gamma'} + e^- + H_2O^+ \quad (8)$$

If the energy level of the recoil photon is sufficient, then this photon can excite an additional water molecule. The whole procedure can be repeated numerous times until the energy level of the recoil photon is sufficiently low, which means that only the Einstein photo-effect occurs.

Usual Roentgen rays with energy $E_{photon} > 15$ keV or $>> 15$ keV and γ-quanta with photon energies $E_{photon}$ in the MeV domain can only lose their energy by Compton interaction. The energy balance is given in eq.( 9).

$$E_\gamma = E_{\gamma'} + E_{electron} \quad (9)$$

In addition to the recoil photon, the released electron due to each Compton process is itself capable to ionize various $H_2O$-molecules. With regard to the Compton process, we should like to list some further important reactions with photons and released electrons according to eq. (8):

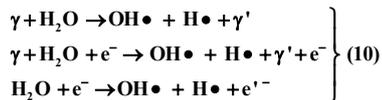

$$\left.\begin{array}{l} \gamma + H_2O \rightarrow OH\bullet + H\bullet + \gamma' \\ \gamma + H_2O + e^- \rightarrow OH\bullet + H\bullet + \gamma' + e^- \\ H_2O + e^- \rightarrow OH\bullet + H\bullet + e'^- \end{array}\right\} \text{ (10)}$$

$H\cdot$ and $OH\cdot$ represent radical states, and $e'^-$ refers to a recoil electron with lower energy compared to the preceding electron $e^-$ before scatter. A variety of possible reactions and interactions can occur during the collision of photons with water molecules, all of which lead to the formation of excited states. These excited states are important intermediates of the overall water splitting reaction (see e. g., eq. 10). Ionizing radiation effectively increases the density of excited states as compared to the absence of ionizing radiation.

## 2.2. Separation of ions produced by γ-irradiation

When ionizing radiation interacts with water, a mixture of positive and negative ions will form via γ-irradiation by a conglomerate of ions. According to the specific radiation beam properties, including the transverse profiles at each depth of the passing beam, the conglomerate shows the shape of a Gaussian distribution resulting from the beam scatter. Conventional rod- or foam-shaped hydrolysis electrodes are inefficient to achieve the task to separate the positive ions from their negative counterpart (ideally, all cations should migrate to the cathode and all anions should migrate to the anode). In order to improve the separation of negative from positive ions, plate-shaped electrodes are used. This electrode shape allows the plates to fulfil a dual function: on one hand, they act as electrodes, on which the $H_2$ and $O_2$ evolution reactions takes place. On the other hand, the cathode and anode act as the negative and positive pole of a *plate capacitor*. A strong electric field builds up between the electrodes, which effectively separates the ions more effectively than rod- or foam-shaped electrodes (this fact could be demonstrated by suitable experiments based on 6 MV radiation beam produced by a linear accelerator, Ulmer et al., 2005).

To amplify, if we assume, in a first step, a Gaussian distribution, then a recombination of the ions would lead to neutral water molecules, again, but this is not the goal of the study. A description of the distribution of ions can be solved via well-known diffusion equations in the succeeding section.

The questions arise, whether the ions migrate to the electrodes. This is true in every case for OH⁻, but every extinction of this ion implies a release of such ions from water in order to maintain the thermodynamic equilibrium under the given conditions according to eq. (4).

## 2.3. Basic diffusion equation including electric field with radiation

The physical process can be described by diffusion of the scattered electrons with the help of the following diffusion equation subjected to an electric field produced by a cathode and anode used by plates at the wall of the aquarium:

$$- \; \partial C/\partial t \; + D\cdot\Delta C + \alpha\cdot x\cdot C = 0 \text{ (11)}$$

Eq. (11) represents the well-known diffusion equation in 3 dimensions extended by an electric field based on methods previously worked out (Ulmer, 1985).
The factor $\alpha$ is given by either $Fe_0\cdot(U/d)/h$ with F: Faraday constant and $e_0$ elementary charge or by $Q\cdot(U/d)/h$ with Q: total charge of the medium produced by radiation. Further abbreviations are D: diffusion constant; U: voltage; d: distance between the plates of the capacitor and *h*:

Planck's constant divided by $2\pi$. The electric field strength results from U/d = |**E**|. The above stated parameters result from the transition of an electric field with ionization densities in the electric field with the Schrödinger equation, if the time-derivative is imaginary, and the substitution $t \rightarrow i \cdot t$ is performed. The diffusion constant D has to satisfy the Einstein-Smoluchowski relation:

$$D = k_B \cdot T/(6\pi \cdot \eta \cdot r) \qquad (12)$$

The parameters of eq. (12) are given by T: absolute temperature, $k_B$: Boltzmann constant, $\eta$: viscosity of the medium and r: ion radius. The following numerical values can be assumed: The viscosity $\eta$ of $H_2O$ is $\eta = 1$ at 20° C and $\eta = 0.891$ at 25° C. The ion radius r of H (proton) is $r = 0.85 \cdot 10^{-13}$ cm and of Oxygen it amounts to about $r \approx 6 \cdot 10^{-13}$ cm.

The above equation is solved by eq. (13):

$$C = A \cdot exp[(i \cdot (k_1 x + k_2 y + k_3 z)]exp(\gamma t) \cdot exp[(-(\tau_1 t + \tau_2 t^2 + \tau_3 t^3)] \quad (13)$$

The parameters of this equation have to satisfy

$$\alpha = \gamma; \qquad (13a)$$
$$\tau_1 = D \cdot (k_1^2 + k_2^2 + k_3^2); \qquad (13b)$$
$$\tau_2 = i \cdot D \cdot k_1 \cdot \gamma; \; \tau_3 = D \cdot \gamma^2. \qquad (13c)$$

Due to the linearity of the above equation the integration over the vector **k** leads to the integral (14):

$$C = \iiint A \cdot exp[(i \cdot (k_1 x + k_2 y + k_3 z)] \cdot exp(\alpha \cdot t) \cdot exp[-(\tau_1 t + \tau_2 t^2 + \tau_3 t^3)]dk_1 dk_2 dk_3. \quad (14)$$

The solution of this integral is given by:

$$C = N \cdot exp[-((x - D \cdot \alpha \cdot t^2)^2 + y^2 + z^2)/(4 \cdot D \cdot t)] \cdot exp(\alpha \cdot x \cdot t) \cdot exp(-D \cdot \alpha^2 \cdot t^3/3). \quad (15)$$

**The normalization constant N is given**

**by: $1/(4 \cdot \pi \cdot D \cdot t)^{3/2}$. (15a)**

For the reason of symmetry of the capacitor a further solution exists, namely by replacing x by $x' \rightarrow d-x$:

$$C = N \cdot exp[-((x' - D \cdot \alpha \cdot t^2)^2 + y^2 + z^2)/(4Dt)]exp(\alpha \cdot x' \cdot t)exp(-D\alpha^2 t^3/3). \quad (16)$$

The normalization factor N remains unchanged.

Eq. (11) can be solved exactly, but by taking account of some wall restriction given by the electric field producing capacitor, the solution (13c) represents a rather satisfactory solution property.

The resulting ion concentration profiles are plotted in Figure 2 and 3. Figure 2 shows the distribution at the beginning of the experiment, immediately after the ionizing radiation is inserted into the apparatus. Figure 3 shows the ion distribution after the time of about one second.

Both Figures show the effect of an electric field produced by the capacitors in order to separate the ions according to the red curves. As long as the irradiation is present, a stationary state between new production of ions and transport of them to the capacitor plates will be established. A principal difference between the two figures is an accumulation effect, i.e., the travelling ion concentrations are increased due to the comparably slow diffusion velocity. This property changes the relative amplitude.

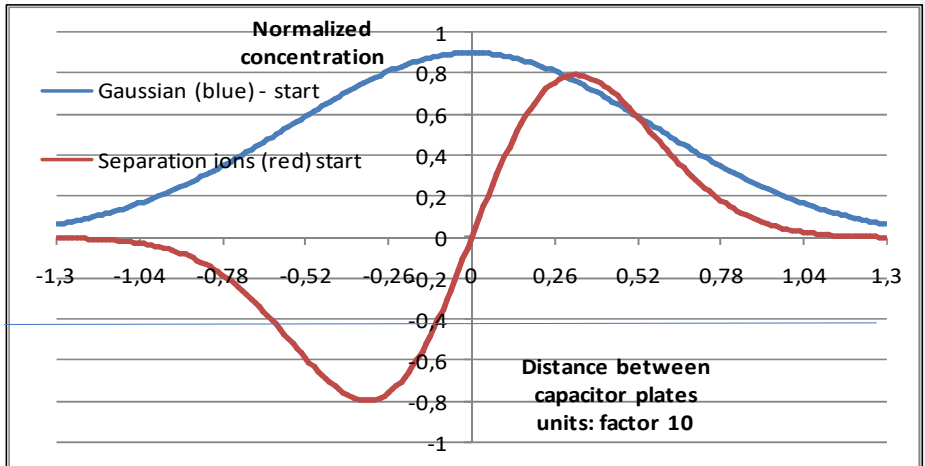

**Figure 2.** At t = 0 the Gaussian can rather describe the ion distribution (blue), curve red shows the polarization at t = 0.

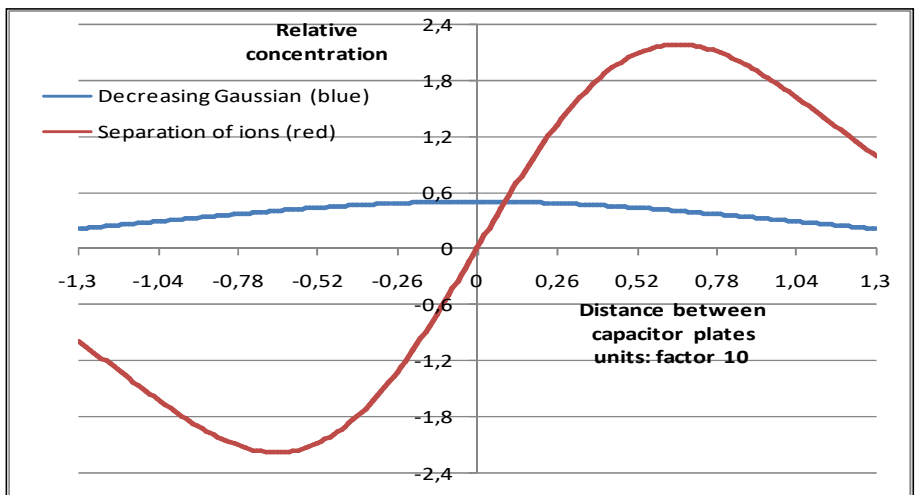

**Figure 3**. The original Gaussian distribution has nearly been vanished, and the positive ions (positive part of the amplitude) travel in direction to the cathode, whereas the negative ions (negative part of the amplitude) travel in direction to the anode.

## 2.4. Monte-Carlo (MC) calculations

The complete task, namely the creation of γ-radiation and the influence of an electric field in accordance with chosen geometrical conditions can be simulated with the well-known MC code GEANT4, which is successfully applied and documented by publications in many disciplines of theoretical and applied physics (see e.g., (Ulmer, 2005 and references therein). In this communication, we have applied this code with regard to the configuration in section 2.6, i.e., the ionization produced by 6 MV in an aquarium and an additional electric field according to Figure 4. The advantage of MC calculations is given by readily modified geometrical alterations, which would require to completely new apparatus constructions.

## 2.5. Experimental performance

Initial calculations were performed to identify the most suitable state (liquid or gas) of the water for the experiment, as well as for a large-scale implementation of the technology. These calculations revealed that liquid water exhibits much higher absorption cross-sections of γ-rays of $Cs^{137}$ than steam. 18 cm of liquid water absorb the same amount of radiation of wavelength of the order $\lambda \approx 2 \cdot pm$, whereas the same absorption in steam requires ca. 40 m.

Test measurements have been carried out at the linear accelerator (Linac). Their purpose was to identify the effect of ionizing radiation on the electrolytic water splitting reaction. This accelerator was chosen, because the wavelength of the radiation is similar to the energy of the radiation produced during the radioactive decay of the isotopes present in the waste of nuclear reactors. The both capacitors consist of high-grad steel with thickness 1 mm, length: 11 cm and height: 12 cm. The solenoids are also of high-grad steel and the wires connected to a power supply are of copper. These means have been purchased by a ware-store; the water box could be purchased from Mollenkopf, GmbH, in Stuttgart in the form of an aquarium in order to get the rapid availability of a test phantom. The power supply with maximum 30 V and 20 Å has been purchased from an electronic marked center.

## 2.6. Measurement apparatus

A simplified schematic of the apparatus used for the measurements is shown in Figure 4.

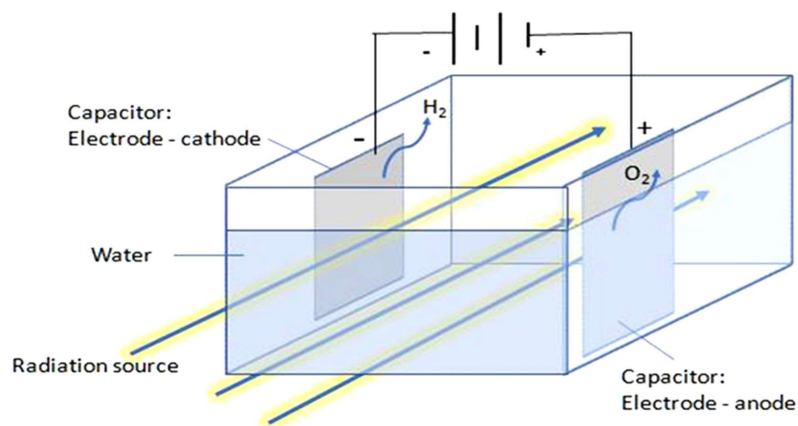

**Figure 4.** Basic configuration of an electrolyser using external γ-rays and two capacitors for cathode and anode
.

A box with dimensions 18 cm x 13 cm x 18 cm was used for the measurements. The box (bottom and walls) consists of special glass (thickness: 5 mm).  Electrodes of various shapes and compositions (rods or plates made from stainless steel) were inserted into the box and connected to DC power supplies. In one set of experiments, the aquarium box was subdivided into two separate compartments by a diaphragm to avoid the formation of explosive gas mixtures according to Figure 5.

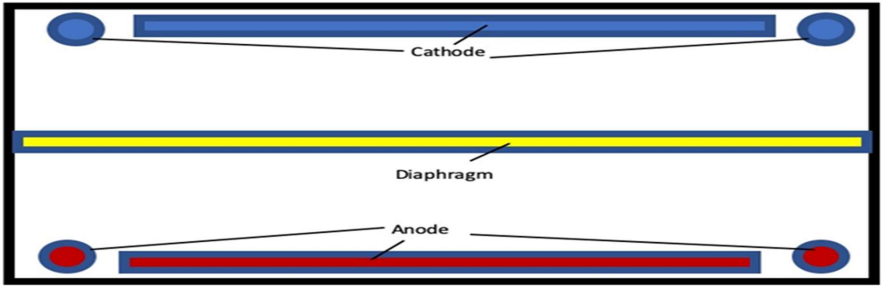

**Figure 5.** Top view on electrolyzer with the capacitor plates and additional electrodes. The use of a diaphragm is optional and serves for separation of $H_2$ and $O_2$ gases to avoid the formation of explosive gas mixtures.

After arrangement of all components, the aquarium box was closed using an acrylic glass-plate and sealed using glue. The acrylic glass top contained several valves, through which water could be introduced into the apparatus and the formed gasses could be released. The apparatus was tested for leakage before each experiment. The amount of gas produced in a given time period was measured by displacing water in an upside-down graduated cylinder immersed in a water tank. Cu-based electrodes were found to be less stable and robust during the experiments. Therefore, stainless-steel electrodes were used for all experiments. For the experiments, distilled water containing ca. 50 g NaOH/2.5 L water was used. were used, Figure 5. Using the depth dose curves of a Linac, the applied dose of a 10 x 10 cm$^2$ field of 6 MV, according to Figure 6. At the beam entrance the dose amounts to 150 %, and at the end of the box the dose decreases to 75 % of the entrance value. Thus, at the depth of 10 cm in the water box 1 Gy was applied.

For the first successful set of measurements, plates of high-grade steel and the two electrodes at each side (both cathode and anode sides) of the plates.

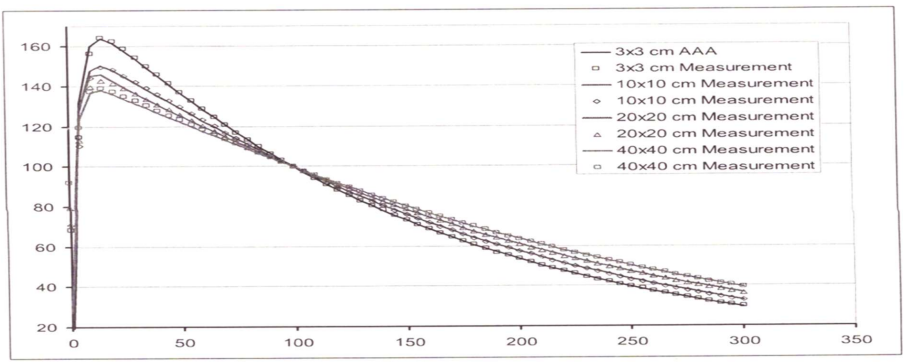

**Figure 6.** Depth dose curves of 6 MV 'bremsstrahlung' of a Linac (Ulmer, et al., 2005).

It is usual in radiation physics to represent the depth of absorption materials (here water) in mm (abscissa) and in the ordinate the relative percentage of dose absorption (normalized to a depth of 100 mm). According to the measurement properties (geometry of the water box) only a beam with a lateral cross-section of 10 x 10 cm$^2$ is applicable. The beam properties show a build-up effect at the surface and a maximum dose at the depth 1,9 cm. If 100 MUs in one minute are applied (1 Gy at depth 10 cm), then at surface the dose amounts to 1,5 Gy and at the end of the box 0,75 Gy. The dose at the entrance surface only amounts to 30 % (0,5 Gy, build-up domain in Figure 7a), since the released electrons travel towards the inside of the aquarium box, where the maximum is reached. The measurements were performed using similar or slightly higher radiation energies as would be released during the radioactive decay of Cs[137], the most abundant isotope in radioactive waste.

## 3. Results

### 3.1. Description of the measurement conditions without radiation

The described water box contained 2,5 L two-fold distillated water and 50 g NaOH as electrolyte.
The outer walls were allocated with 4 solenoid-like electrodes of high-grad steel. The material of the capacitors on both opposite sites were high-grad steel with 2 mm thickness and area of 15 x 10 cm$^2$. Both, capacitor and solenoids, were supplied with two independent voltage suppliers. It could be verified that a mutual influence existed, i.e., current and voltage could not be controlled of one system without changing the parameters of the other one.

### 3.2. Measurements with capacitor and electrodes and results

The abbreviations in Table 1 and Table 2 are:
E: Electrodes; C: Capacitor; Sole: either only capacitor or electrodes are involved at measurement process; Both, Capacitor and electrodes are involved, Dose in Gy, MU: monitor units per minute. The results are summarized in Figures 7.1 and 7.2. Figure 7.3 represents a multi regression analysis of the measured data, which are the basis for the Monte-Carlo calculations.

**Table 1.** Results of the first series without radiation

| Voltage/V | Current/A | Configuration | Exposure time H$_2$O/mL | Displacement H$_2$O/mL |
|-----------|-----------|---------------|-------------------------|------------------------|
| 3,7 | 0,79 | E | B | B |
| 3,8 | 1,48 | C | 8 min | 205 |
| 3,5 | 0,34 | E | B | B |
| 5,0 | 2,79 | C | 5 Min | 210 |
| 5,9 | 1,28 | E | B | B |
| 8,1 | 5,16 | C | 3 Min | 210 |
| 4,3 | 0,33 | E | B | B |
| 7,4 | 5,16 | C | 3 Min 30 sec | 200 |
| 3,2 | 0,82 | E (S) | 8 Min | 85 |
| 3,4 | 1,34 | C (S) | 9 Min | 90 |
| 7,5 | 5,16 | C  (S) | 2 Min | 200 |

The first measurements with radiation carried out at a Linac in the ‚Strahlenzentrum am Maximiliansplatz 2, Munich‘ has been performed with the following conditions:
Identical experimental conditions for electric supply, concentration of NaOH and two-fold distillated water. Radiation conditions at the Linac: SSD: 89,5 cm; field size et entrance: 10 x 10 cm$^2$, 6 MV. Dose at entrance: 1,28 Gy and at the end of the water box: 0,7 Gy. The irradiation was performed with three different intensities: 1 Gy (100 monitor units: MU) in one minute, 200 MU in 30 seconds, 3 MU in 20 seconds. The dose intensity in the production of 'Bremsstrahlung' of a Linac is free to be chosen.

All measurement series refer to a field size of 10 x 10 cm² and 6 MV. Further details can be taken from the protocol. There are two particular features: The yield of Hydrogen reached the highest value, when the radiation intensity amounted to only 100 MU/min. The augmentation of the radiation intensity to 200 MU/min or 300 MU/min did not deliver the factor two or three with reference to the yield of Hydrogen. The most outstanding result was obtained by a factor of about 9 – 10, when only 100 MU/min two minutes have been exposed, and the identical voltage and current have been used for comparison with absence of radiation (this is the so-called zero experiment).

What can we conclude from these findings? An increase of the radiation intensity requires an increase of the voltage of the power supply in order to attract the ions to the cathode/anode. The four electrodes did only show to have a local influence, since the results at the case with radiation were nearly identical, when only the capacitor plates were connected to the power supply.

**Table 2.** Results of the series with radiation

| Volt/V | Current/A | Configuration | Exposure time and MUs/Min | Displacement H₂O/mL |
|---|---|---|---|---|
| 8,2 | 5,17 | E | 1 Min 10 s | B |
| 8,2 | 6,18 | C | 100 MUs | 260 |
| 7,4 | 5,16 | E | 1 Min 15 s | B |
| 8,2 | 5,16 | C | 100 MUs | 245 |
| 7,4 | 6,80 | C | 1 Min 30 s | B |
| 8, | 5,16 | E | 100 MUs | 330 |
| 7,4 | 6,76 | E | 90 s | B |
| 8,2 | 5,16 | c | 100 MUs, Dose: 300 G/Min | 290 |
| 7,4 | 6,73 | E | 100 MUs Dose: 3 Gy/Min | B |
| 8,2 | 5,16 | C | 1 Min 28 s | 362 |
| 8,2 | 5,16 | C (S) | 100 MUs Dose: 1 Gy/Min | 310 |
| 8,2 | | C (S) | 3 Gy/Min | |
| 8,0 | 5,16 | C(S) 1 Min | 100 MUs Dose: 1 Gy/Min | 120 |
| 3,8 | 1,28 | E (S) | 100 MUs 2 Min 10 s | 20/25 |
| 4,1 | 1.31 | K (S) | 100 MUs 2 Min 12 s | 195 |

| Configuration | Voltage/V | Current/A | MU/related to 1 minute | Total dose in Gy | Displacement per minute: mL / D$_L$ |
|---|---|---|---|---|---|
| C | 4.02 | 1.295 | 100 | 2 | 97 (S) |
| E | 4.02 | 1.295 | 100 | 2 | 11 (S) |
| C | 4.02 | 1.295 | 0 | 0 | 10 (B) |
| E | 4.02 | 1.295 | 0 | 0 | 10 (B) |
| C | 8.02 | 6.82 | 100 | 1 | 445 (B) |
| E | 7.41 | 5.19 | 100 | 1 | 445(B) |
| C | 8.02 | 6.82 | 0 | 0 | 66 (B) |
| E | 7.41 | 5,19 | 0 | 0 | 66 (B) |
| C | 7.42 | 2.96 | 0 | 0 | 55(B) |
| E | 7.42 | 2.96 | 0 | 0 | 55(B) |
| C | 7.42 | 2.96 | 100 | 1 | 336 (B) |

| | Volt/V | Current/A | MU/Min | Dose/Gy | Displacement mL |
|---|---|---|---|---|---|
| E | 7.42 | 2.96 | 100 | 1 | 336 (B) |
| C | 7.42 | 2.96 | 0 | 0 | 55(B) |
| E | 7.42 | 2.96 | 0 | 0 | 55(B) |
| C | 7.42 | 2.96 | 100 | 2 | 320 (B) |
| E | 7.42 | 2.96 | 100 | 2 | 332 (B) |
| C | 7.42 | 2.96 | 0 | 0 | 55(B) |
| E | 7.42 | 2.96 | 0 | 0 | 55(B) |
| C | 7.42 | 2.96 | 100 | 3 | 325 (B) |
| E | 7.42 | 2.96 | 100 | 3 | 325 (B) |

**Table 3.** Some examples of measurement conditions.

| | Volt/V | Current/A | MU/Min | Dose/Gy | Displacement mL | Time |
|---|---|---|---|---|---|---|
| C | 4,02 | 1,295 | 100 | 2 | 97 | 2 Min |
| E | 4,02 | 1,295 | 100 | | 11 | 2 Min |
| C | 8,02 | 6,82 | 100 | 1 | 445 | 1 Min |
| E | 7,41 | 5,19 | 100 | 1 | 445 | 1 Min |
| E+C | 8,02 | 6,82 | 0 | 0 | 66 | 3 Min |
| E+C | 7,41 | 5,19 | 0 | 0 | 66 | 3 Min |

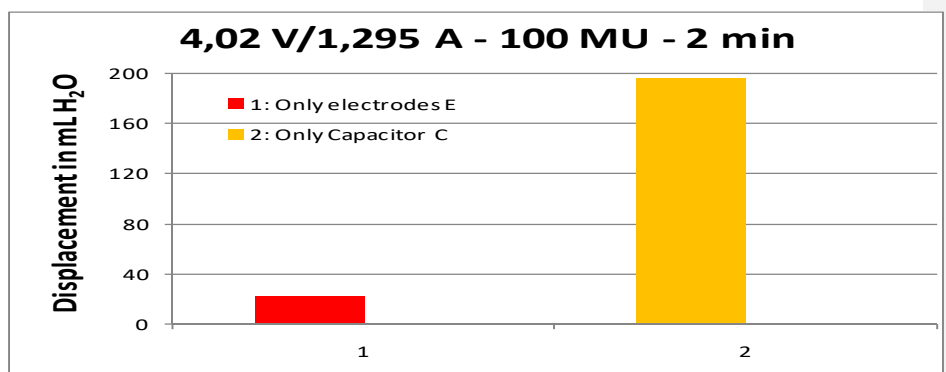

**Figure 7.1.** Comparison 100 MU: 1 Gy only electrodes (red) and only capacitor (yellow) - this represents the average of one measurement repeated 3 times under the conditions stated in the Figure. The statistical deviations amounted to 1.2 %.

The multi-regression analysis of the measurement series with and without radiation is present in Figures 7.2 and 7.3. If the measurement is performed only by one modality of voltage supply (either capacitor or electrodes) then the index S is used! If the measurement is performed with both modalities, then the index B is used. All results are uniquely normalized to a one-minute exposure time with and without radiation!

Dose rate $D_R$: Number of MUs per minute. The value had to be corrected, because the Linac required always 30 seconds more! Gy: The applied dose is always 1 Gy at depth 10 cm (discussed in detail in the description. This statement refers to the delivered total dose! Explanation: If MU = 0 and dose = 0 then measurement without irradiation.

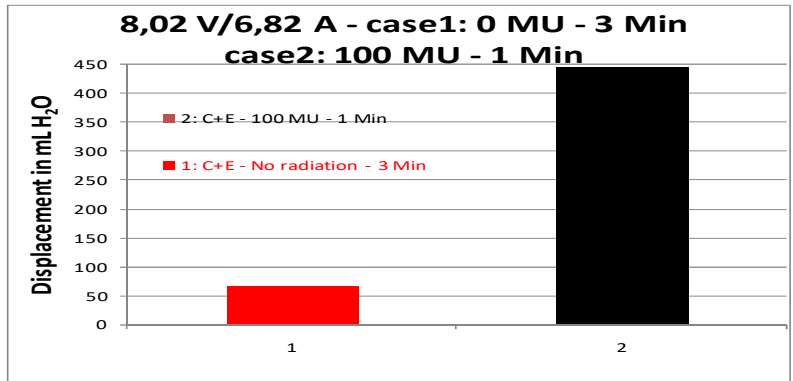

Figure 7.2. Multi-regression analysis - a comparison case 1 (no radiation) and case 2 (radiation).

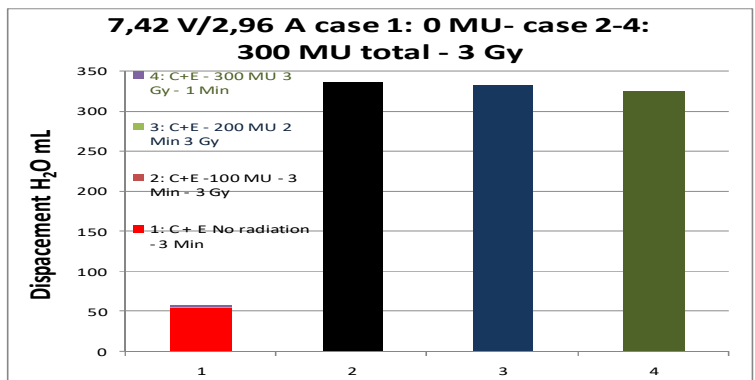

**Figure 7.3.** Multi-regression analysis: Comparison case 1 no radiation with cases 2 - 4, radiation with different dose rates, total Dose: 3 Gy.

### 3.3. Monte-Carlo (MC) calculations with GEANT4

As already pointed, the MC calculations have to account for the geometrical boundaries according to Figure 4 in chapter 2.6, which are given by the propagation of the external radiation in z-direction and the arrangement of the capacitors in perpendicular direction. In order to verify the reliability of the multi-regression analysis, we have used the optimization results according to Figures 7.2. and 7.3. If one regards the depth-dose curve of 6 MV according to Figure 6, one should assume that the output of dose does not have an influence. However, this is not quite correct, since the movability of the ions under an electric field seems to play a certain role, but under the experimental conditions according to the above Figure 7.3 the influence is only of minor importance. Therefore, the MC calculations did not account for this difference. Thus, the deviation in Figure 7..2. amounted to +1.6 % (red case, without irradiation) and +1.9 % (black case, with radiation). In Figure 7.3 we obtained + 1.7 % (red case, without radiation) and + 1.8 % (average of the three cases).

It is noteworthy to mention the modification of the box geometry (material: Lucite instead of aquarium glass) by reduction of the phantom thickness to 6 cm, the capacitor plates are identical and located at the phantom walls, the current and the voltage were reduced to the half compared with the aquarium conditions. By that, the Hydrogen outcome increased by 12 % (red case) and by 13.4 % at the cases with radiation. This difference clearly shows that the mobility of the ions can be exploited in a more suitable way. The commitment of MC calculation appears to represent a useful toolkit in order to reduce experimental trials.

# 4. Conclusions and outlook into the future

## 4.1. Conclusions

These results demonstrate the essential role of ion creation by the application of ionizing radiation. A look at Figures 2 and 3 already provides an indication that via reduction of the currents, voltages and the thickness of a phantom this exploitation does not behave in a linear fashion in dependence of the current, voltage and phantom thickness, since all three parameters only show approximately a linear behavior according to the solution function (eq. 15) of the diffusion (eq. 11) extended by an additional electric field. With regard to the use of radioactive waste one has to account for this behavior. The account for the practical feasibility indicates that the phantom thickness only should amount to 5 cm – 6 cm, and the phantom length should be restricted to about 10 cm – 12 cm, if opposite sources consisting mainly of $Cs^{137}$ are taken into consideration. This conclusion is supported by the MC calculations, and this aspect is also considered in the succeeding sections. A high number of potential improvements and modifications to the present apparatus are possible. These are sketched below.

# 4.2. Some feasible extensions for technical applications

## 4.2.1. Diaphragm

The diaphragm, which separates the anode from the cathode compartments, is necessary in order to separate $H_2$ from $O_2$. However, the solution used in these experiments can significantly be improved by semi-permeable walls or layers. Another possibility to reach this separation might be specific layers at the non-return valves.

## 4.2.2. Electrolytes (possibly irrelevant in the presence of radiation)

The reported electrolyte for the test measurement was NaOH. In general, bases and $H_2SO_4$ may be candidates as electrolytes. If $H_2SO_4$ is considered, then it is necessary that the inner part of the box is equipped with a thin noble metal plate, which is resistant against this acid. HCl must be excluded, because the production of $Cl_2$ is not desired, and $HNO_3$ would, besides $H_2$, also release $NH_3$. The bases NaOH, KOH, $Mg(OH)_2$, and some other bases in weak or medium concentration are suitable candidates. The effect of these electrolytes on the performance of the cells must be evaluated in the future.

## 4.2.3. Capacitor plates

Present experiments have been performed using high-grade steel alloys. The use of copper for long-run applications may be possible, if the electrolytes cannot undergo reactions under radiation conditions. Due to the release of ions in water by radiation exposure, it is possible to avoid electrolytes at all, this property would treat the material with more care. The use of noble metals as capacitor plates is possible, but only a question of costs. Very interesting capacitor plates positioned at the walls are of tungsten or tantalum, but the plates must be isolated from the walls, if the wall material is chosen to be not an insulator. It may be beneficial to use structured plates to increase the active surface area, or deposit co-catalysts on the plates.

### 4.2.4. Walls of the Aquarium Boxes: Replacement by Lucite

From radiotherapy with protons/photons is known that Lucite is also a very resistant material against ionizing radiation; it is much easier to manipulate mechanically than the Laboratory glass of the used test phantom. The aquarium glass used for phantoms of the tests was easy to purchase, but for technical applications it is not suitable.

### 4.2.5. Effect of an inhomogeneous magnetic (static) field

It is well-known that inhomogeneous magnetic fields direct charged particles either to positive or negative directions of their motion. Thus, the poles have to be chosen in a proper way to enhance the transport of $H^+$ ions to cathode and $OH^-$ ions to anode. An additional influence results from the magnetic interaction with the spin of the particles, and, above all, $O_2$ is affected by such a field due to its triplet state. This is, however, a second order effect.

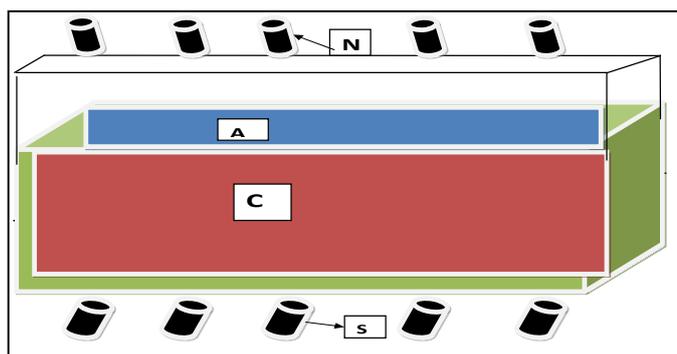

**Figure 8.** An inhomogeneous magnetic field can enhance the yield of $H_2$. The magnets with north poles **N** and south poles **S** must possess a bigger extension as show in the above Figure. All magnets are permanent magnets (ferro-magnets). The radiation beam passes between these plates.

### 4.2.6. Use of radioactive decay products of nuclear reactors

We have to point out that the goal of this study is the enormous source of radiation energy resulting as decay products (radioactive waste) of nuclear reactors after separation of $U_{235}$ and Plutonium, which both are made again applicable for new fission processes (this procedure is performed in a so-called reprocessing plant). Relations (9, and 10) show possible nuclear reactions of $Cs^{137}$ irradiated with protons. Thus eq. (9) would provide a rather nice reaction product, namely $Cs^{136}$ with a half-time of about 22 days (Etspüler, 2012, Kulke, 2010, Ulmer, 2016). The second possible reaction referred to as equation (10) with protons delivers the element Barium, which is a stable metal. Thus, the energy of the proton must not exceed 70 MeV in order to enable the above nuclear reactions and to prevent other undesired nuclear processes. The relationship between the required proton energy and the released γ-quanta is also an open question, because the principal way how to exploit the γ-energy according to the above equations 9 and 10 is also not yet elaborated. Nevertheless, it may be possible and useful to combine a transmutation technology with the method of enhanced Hydrogen production by exploiting the γ-radiation of radioactive decay products of nuclear reactors. With regard to nuclear reactions occurring in reactors it should be recalled that a nuclear fission induced by a neutron collision with $U^{235}$ (or a similar nucleus such as Plutonium) usually produces two fission nuclei with about the half mass of the origin nucleus. Therefore, the most important fission products are the isotopes:

Cesium, Barium, Strontium, but also $Co^{60}$ and Iodine. Thus, only Iodine isotopes exhibit a rather short half-time, and, by that, they are not a problem with regard to radioactive waste.

As already pointed out, $Cs^{137}$ represents the most outstanding fission product.

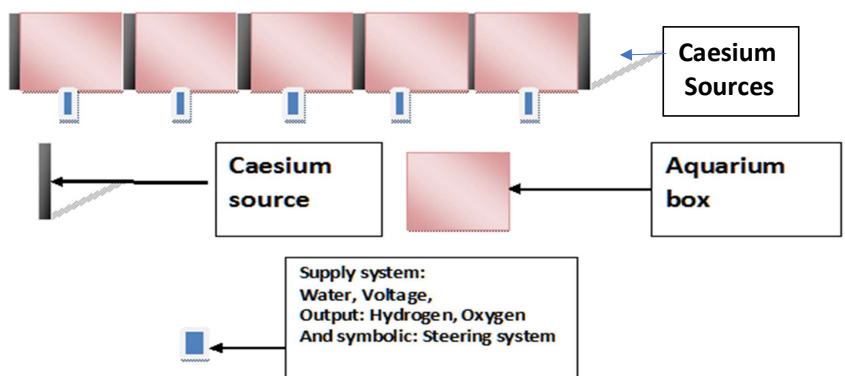

**Figure 9**. Model of a sequence of phantom cells for the production of $H_2$ and $O_2$.

Several boxes consisting of Lucite would be placed in series. They could be separated As pointed out the γ-energy of $Cs^{137}$ amounts to 0,7 MeV. This fact implies that a water box with the length 18 cm is irradiated with a nearly homogeneous dose distribution, if the box is irradiation by an opposite field (one slice of $Cs^{137}$ on each end of the aquarium box). The appropriate voltage depends on the activity and can be reduced in due course. By a measurement and comparison with the Linac results one is able to find the optimum of the voltage. The thickness of the $Cs^{137}$ slices may be of the order of 2 mm, placed between the aquarium boxes. The complete arrangement is shown in Figure 8, but the number of aquarium boxes is not restricted by the drawing and may enormously be stretched. Thus about 100 cells in one sequence require a length of ca. 18 m – 19 m. It is also obvious that such an arrangement of 'cells' for the gas production requires remote and control implementations in order to survey all functions and to satisfy the radiation protection.

## Acknowledgment

Many thanks go to U. Wild, Radiation Center Munich for making the tests at the Linac possible and U. Ulmer for experimental support.

Kommentiert [w1]: